\documentclass[prd,onecolumn,nofootinbib]{revtex4}
\usepackage{bm,amsmath,amssymb,graphicx}

\begin{document}

\title{Spinors with torsion and matter--antimatter asymmetry}
\author{Nikodem Pop{\l}awski}

\altaffiliation{NPoplawski@newhaven.edu}
\affiliation{Department of Mathematics and Physics, University of New Haven, West Haven, CT, USA}

\begin{abstract}
The conservation law for the orbital plus spin angular momentum of a free Dirac particle in curved spacetime requires that the affine connection has the antisymmetric part: the torsion tensor, which extends general relativity to the Einstein--Cartan theory of gravity.
In the presence of torsion, the Dirac equation becomes a nonlinear, cubic equation in the spinor wave function.
We show that the energy eigenvalues of the corresponding Hamiltonian as functions of the momentum are different for the fermion and antifermion components of the spinor, violating charge conjugation symmetry, and also depend on the helicity.
Consequently, particles of matter and antimatter have different dispersion relations and therefore different masses.
This mass difference increases with density and becomes significant near the Cartan density, which existed in the early Universe.
Because antimatter particles were more massive than matter particles, they were also slower during pair production in the early Universe and therefore had higher cross sections for gravitational capture by primordial black holes.
This difference might have led to the matter--antimatter imbalance in the observable Universe: the missing antimatter fell into black holes.
\end{abstract}
\maketitle

\noindent
{\bf General relativity with spin and torsion}.\\
The Einstein--Cartan theory is the simplest and most natural theory of gravitation that extends general relativity to include spin, by removing the symmetry condition of the affine connection $\Gamma^{k}_{ij}$.
The antisymmetric part of the affine connection: the torsion tensor $S^k_{\phantom{k}ij}=(1/2)(\Gamma^{k}_{ij}-\Gamma^{k}_{ji})$, is produced by the spin angular momentum of matter \cite{EC}.
In this theory, the Lagrangian density for the gravitational field is given by the Ricci scalar and the covariant derivative of the metric tensor is zero, as in general relativity.
The connection is given by $\Gamma^{k}_{ij}=\mathring{\Gamma}^{k}_{ij}+S^k_{\phantom{k}ij}+S_{ij}^{\phantom{ij}k}+S_{ji}^{\phantom{ji}k}$, where $\mathring{\Gamma}^{k}_{ij}$ are the Christoffel symbols.
The field equations are obtained from the variations of the total action for the matter and gravitational field with respect to the metric and torsion tensors, giving the Einstein equations and the Cartan equations $S^i_{\phantom{i}\mu\nu}-S_\mu e^i_\nu+S_\nu e^i_\mu=-(\kappa/2)s^{\phantom{\mu\nu}i}_{\mu\nu}$, where $s^{\phantom{\mu\nu}i}_{\mu\nu}$ is the spin tensor, $e^i_\mu$ is the tetrad with Latin vector indices and Greek Lorentz indices, and $\kappa=8\pi G/c^4$.
We use the notation of \cite{FranciscoNiko}.\\

\noindent
{\bf Dirac equation with torsion}.\\
The covariant derivative of a spinor $\psi$ is given by $\nabla_i\psi=\partial_i\psi-\Gamma_i \psi$, where the spinor connection $\Gamma_i$ is given by the Fock--Ivanenko coefficients 
$\Gamma_i=-(1/4)\omega_{\mu\nu i}\gamma^\mu \gamma^\nu$, with the spin connection $\omega^\mu_{\phantom{\mu}\nu i}=e^\mu_k\omega^k_{\phantom{k}\nu i}=e^\mu_k \nabla_i e^k_\nu=e^\mu_k(\partial_i e^k_\nu+\Gamma^{k}_{ji}e^j_\nu)$ and the Dirac matrices $\gamma^\mu$ \cite{FranciscoNiko}.
The Dirac equation for a spinor wave function representing a fermion with mass $m$ is given by $i\hbar \gamma^\mu e_\mu^i \nabla_i \psi = mc\psi$.
In the Einstein--Cartan theory, the covariant derivative in the Dirac equation is linear in torsion, the torsion tensor is proportional to the spin tensor because of the Cartan equations, and the spin tensor is quadratic in the spinor field and completely antisymmetric: $s^{\mu\nu\rho}=(1/2)i\hbar\bar{\psi}\gamma^{[\mu}\gamma^\nu\gamma^{\rho]}\psi$, where $\bar{\psi}=\psi^\dag\gamma^0$ is the adjoint spinor.
Consequently, the Dirac equation is cubic (nonlinear) in spinor fields \cite{EC,FranciscoNiko}:
\begin{equation}
    i\hbar \gamma^\mu e_\mu^i\mathring{\nabla}_i\psi + \frac{3}{8}\kappa\hbar^2(\bar{\psi}\gamma_\mu\gamma^5\psi)\gamma^\mu\gamma^5\psi = mc\psi,
    \label{Dirac}
\end{equation}
where $\mathring{\nabla}_i$ is the covariant derivative with respect to $\mathring{\Gamma}^{k}_{ij}$.

In the absence of spin, torsion vanishes, so the affine connection reduces to the Christoffel symbols and the Einstein--Cartan theory reduces to general relativity \cite{GR,LL2}.
This theory is also indistinguishable in predictions from general relativity at densities of matter much lower than the Cartan density $\rho_\textrm{C}=m^2 c^4/G\hbar^2$ \cite{non}, which is about $10^{51}$ kg/m$^3$ for the electron, so it passes all observational and experimental tests of general relativity \cite{EC}.\\

\noindent
{\bf Dirac Hamiltonian matrix with torsion}.\\
In the locally flat coordinate system, the Dirac equation (\ref{Dirac}) can be written in the Hamiltonian form:
\begin{equation}
    \frac{i\hbar}{c}\frac{\partial\psi}{\partial t}=-i\hbar{\bm\alpha}\cdot{\bm\nabla}\psi+mc\beta\psi + \frac{3}{8}\kappa\hbar^2(\bar{\psi}\gamma_\mu\gamma^5\psi)\beta\gamma^\mu\gamma^5\psi,
    \label{wave}
\end{equation}
where ${\bm\alpha}$ is the vector formed from the matrices $\alpha^\mu=\beta\gamma^\mu$ with the space indices $\mu$, and $\beta=\gamma^0$.
If the particle has four-momentum $p_\mu$, the corresponding spinor wave function has a form of a locally plane wave proportional to $\exp(-ip_\mu x^\mu/\hbar)$, where $x^\mu$ are the spacetime coordinates \cite{QM}.
Consequently, the Dirac equation (\ref{wave}) becomes
\begin{equation}
    E\psi={\bf p}\cdot{\bm\alpha}c\psi+mc^2\beta\psi+k(\bar{\psi}\gamma_\mu\gamma^5\psi)\beta\gamma^\mu\gamma^5\psi=H\psi,
    \label{plane}
\end{equation}
where ${\bf p}$ is the momentum of the particle, $E$ is its energy, and
\begin{equation}
    k=\frac{3}{8}\kappa\hbar^2 c.
\end{equation}
Following this equation, the dimension of the wave function is m$^{-1}$ s$^{-1/2}$.

Writing the Dirac matrices in terms of the Pauli matrices $\sigma^\mu$, forming the vector ${\bm\sigma}$, gives the Hamiltonian matrix:
\begin{equation}
    H=\left( \begin{array}{cc}
    mc^2 I_2+k(\psi^\dagger{\bm\alpha}\gamma^5\psi)\cdot{\bm\sigma} & {\bm\sigma}\cdot{\bf p}c-k\psi^\dagger\gamma^5\psi \\
    {\bm\sigma}\cdot{\bf p}c-k\psi^\dagger\gamma^5\psi & -mc^2 I_2+k(\psi^\dagger{\bm\alpha}\gamma^5\psi)\cdot{\bm\sigma}
    \end{array} \right),
\end{equation}
where $I_2$ is the two-dimensional unit matrix, which is equivalent to
\begin{equation}
    H=\left( \begin{array}{cccc}
    mc^2+k\psi^\dagger\alpha_z\gamma^5\psi & k\psi^\dagger(\alpha_x-i\alpha_y)\gamma^5\psi & p_z c-k\psi^\dagger\gamma^5\psi & (p_x-ip_y)c \\
    k\psi^\dagger(\alpha_x+i\alpha_y)\gamma^5\psi & mc^2-k\psi^\dagger\alpha_z\gamma^5\psi & (p_x+ip_y)c & -p_z c-k\psi^\dagger\gamma^5\psi \\
    p_z c-k\psi^\dagger\gamma^5\psi & (p_x-ip_y)c & -mc^2+k\psi^\dagger\alpha_z\gamma^5\psi & k\psi^\dagger(\alpha_x-i\alpha_y)\gamma^5\psi \\
    (p_x+ip_y)c & -p_z c-k\psi^\dagger\gamma^5\psi & k\psi^\dagger(\alpha_x+i\alpha_y)\gamma^5\psi & -mc^2-k\psi^\dagger\alpha_z\gamma^5\psi
    \end{array} \right).
\end{equation}
In terms of the spin pseudovector four-current $j^i=\bar{\psi}\gamma^i\gamma^5\psi$, whose time component is $\rho=j^0=\psi^\dagger\gamma^5\psi$ and space components ${\bf j}$ are $j_x=\psi^\dagger\alpha_x\gamma^5\psi$ and similarly for $y$ and $z$, this matrix is
\begin{equation}
    H=\left( \begin{array}{cccc}
    mc^2+kj_z & k(j_x-ij_y) & p_z c-k\rho & (p_x-ip_y)c \\
    k(j_x+ij_y) & mc^2-kj_z & (p_x+ip_y)c & -p_z c-k\rho \\
    p_z c-k\rho & (p_x-ip_y)c & -mc^2+kj_z & k(j_x-ij_y) \\
    (p_x+ip_y)c & -p_z c-k\rho & k(j_x+ij_y) & -mc^2-kj_z
    \end{array} \right).
    \label{matrix}
\end{equation}

\noindent
{\bf Energy of a free spinor particle with torsion}.\\
Diagonalizing the Hermitian matrix (\ref{matrix}) determines the energy eigenvalues $\lambda$:
\begin{equation}
    \det(H-\lambda I)=\det\left( \begin{array}{cccc}
    mc^2+kj_z-\lambda & k(j_x-ij_y) & p_z c-k\rho & (p_x-ip_y)c \\
    k(j_x+ij_y) & mc^2-kj_z-\lambda & (p_x+ip_y)c & -p_z c-k\rho \\
    p_z c-k\rho & (p_x-ip_y)c & -mc^2+kj_z-\lambda & k(j_x-ij_y) \\
    (p_x+ip_y)c & -p_z c-k\rho & k(j_x+ij_y) & -mc^2-kj_z-\lambda
    \end{array} \right)=0,
\end{equation}
which gives
\begin{eqnarray}
    & & \lambda^4-2\lambda^2[(mc^2)^2+{\bf p}^2 c^2+k^2{\bf j}^2+k^2\rho^2]+8\lambda(k^2\rho c)({\bf p}\cdot{\bf j}) \nonumber \\
    & & +(mc^2)^4+2(mc^2)^2({\bf p}^2 c^2-k^2{\bf j}^2+k^2\rho^2)+{\bf p}^4 c^4+k^4{\bf j}^4+k^4 \rho^4+2k^2 c^2{\bf p}^2 {\bf j}^2-4k^2 c^2({\bf p}\cdot{\bf j})^2 \nonumber \\
    & & -2k^2 c^2{\bf p}^2\rho^2-2k^4{\bf j}^2\rho^2=0.
    \label{dispersion}
\end{eqnarray}
This equation is a quartic equation for $\lambda$:
\begin{equation}
    \lambda^4+a\lambda^2+b\lambda+d=0,
    \label{quartic}
\end{equation}
where
\begin{eqnarray}
    & & a = -2[(mc^2)^2+{\bf p}^2 c^2+k^2{\bf j}^2+k^2\rho^2], \nonumber \\
    & & b = 8k^2\rho c({\bf p}\cdot{\bf j}), \nonumber \\
    & & d = (mc^2)^4+2(mc^2)^2({\bf p}^2 c^2-k^2{\bf j}^2+k^2\rho^2)+{\bf p}^4 c^4+k^4{\bf j}^4+k^4 \rho^4+2k^2 c^2{\bf p}^2 {\bf j}^2-4k^2 c^2({\bf p}\cdot{\bf j})^2 \nonumber \\
    & & -2k^2 c^2{\bf p}^2\rho^2-2k^4{\bf j}^2\rho^2.
    \label{coefficients}
\end{eqnarray}
The solution of this equation determines the four possible values of the energy of a free spinor particle in the presence of torsion.\\

\noindent
{\bf Energy eigenvalues with torsion}.\\
In the presence of torsion, the quartic equation (\ref{quartic}) has four different roots.
They can be determined using the Ferrari method, which writes this equation in the form:
\begin{equation}
    (\lambda^2+y/2)^2=(y-a)\lambda^2-b\lambda+y^2/4-d,
\end{equation}
and chooses the quantity $y$ such that the right side of this equation is a perfect square.
This condition means that the discriminant of the quadratic equation on the right side is equal to zero, giving a cubic equation for $y$:
\begin{equation}
    b^2-(y^2-4d)(y-a)=0.
    \label{cubic}
\end{equation}
For this choice, the quartic equation becomes
\begin{equation}
    (\lambda^2+y/2)^2=(y-a)[\lambda-b/2(y-a)]^2,
\end{equation}
and its roots are given by
\begin{equation}
    \lambda^2-(y-a)^{1/2}\lambda+y/2+\frac{by}{4(y-a)^{1/2}}=0,\quad \lambda^2+(y-a)^{1/2}\lambda+y/2-\frac{by}{4(y-a)^{1/2}}=0.
    \label{energy}
\end{equation}

The cubic equation (\ref{cubic}) can be written as a depressed cubic equation:
\begin{equation}
    t^3+pt+q=0,
    \label{depressed}
\end{equation}
where
\begin{equation}
    t=y-a/3,\quad p=-(a^2/3+4d),\quad q=8ad/3-b^2-2a^3/27,
\end{equation}
and has three different roots.
They can be determined using the del Ferro--Tartaglia--Cardano method, which puts two quantities $u$ and $v$ satisfying
\begin{equation}
    u+v=t,\quad uv=-p/3
\end{equation}
in (\ref{depressed}), giving
\begin{equation}
    u^3+v^3+(u+v)(p+3uv)+q=u^3+v^3+q=0.
\end{equation}
This equation becomes $u^3-p^3/27u^3+q=0$, which with $w=u^3$ reduces to a quadratic equation
\begin{equation}
    w^2+qw-p^3/27=0.
\end{equation}
Solving for $w\to u\to v\to t\to y$ gives $y$ (any of the three roots suffices), which with (\ref{energy}) determines the four energy eigenvalues $\lambda$.\\

\noindent
{\bf Relation to equations of motion with spin and torsion}.\\
In the absence of spin and torsion, the terms with $k$ in (\ref{dispersion}) are omitted and this equation reduces to
\begin{equation}
    [\lambda^2-(mc^2)^2-({\bf p}c)^2]^2=0,
\end{equation}
giving two double roots $\lambda=E$ (positive-energy states describing matter), where $E=[(mc^2)^2+({\bf p}c)^2]^{1/2}$, and two double roots $\lambda=-E$ (negative-energy states describing antimatter).
These roots are consistent with the relation between the four-momentum $P^i$ and four-velocity $u^i$: $P^i=mcu^i$.
In the presence of spin and torsion, the energy as a function of the momentum is given by four different, more complicated values $\lambda$ (\ref{energy}), which are consistent with the relation
\begin{equation}
    P^i=mcu^i+\frac{DS^{ik}}{ds}u_k,
    \label{proportionality}
\end{equation}
where $S^{ik}$ is the four-spin, $D$ is the covariant differential for the affine connection $\Gamma^{k}_{ij}$, and $s$ is the interval.
This relation completes the Mathisson--Papapetrou equations of motion of particles with spin \cite{FranciscoNiko,MP}:
\begin{equation}
    \frac{DS^{ij}}{ds}=P^i u^j-P^j u^i,\quad \frac{DP^i}{ds}=2S_{jk}^{\phantom{jk}i}P^j u^k+\frac{1}{2}R_{jkl}^{\phantom{jkl}i}S^{jk}u^l,
    \label{motion}
\end{equation}
where $R^i_{\phantom{i}klm}$ is the curvature tensor constructed from the affine connection $\Gamma^{k}_{ij}$.

The condition that all four energy eigenvalues are real puts a constraint on the maximum value of $\psi^\dagger\psi$, giving an upper limit on the mass density in the Universe.
This result is consistent with torsion acting at extremely high densities as a repulsive gravitational force that prevents the formation of gravitational singularities.\\

\noindent
{\bf Spinor particle at rest with torsion}.\\
The character of the four different energy eigenvalues for a free spinor particle can be determined by considering the particle in the rest frame of reference, in which the momentum of the particle is equal to zero.
Putting ${\bf p}=0$ and $\rho=0$, which follows from $p_i j^i=0$, in (\ref{coefficients}) gives
\begin{equation}
    a = -2[(mc^2)^2+k^2{\bf j}^2], \quad b = 0, \quad d = [(mc^2)^2-k^2{\bf j}^2]^2.
\end{equation}
The cubic equation (\ref{cubic}) therefore reduces to $y^2=4d$, giving $y=2[(mc^2)^2-k^2{\bf j}^2]$, so $y-a=4(mc^2)^2$.
The solutions (\ref{energy}) of the quartic equation (\ref{quartic}) reduce to $\lambda^2\pm 2mc^2\lambda+(mc^2)^2-(k{\bf j})^2=0$, or $(\lambda\pm mc^2)^2=(k{\bf j})^2$, giving
\begin{equation}
    \lambda=mc^2-k|{\bf j}|,\quad \lambda=mc^2+k|{\bf j}|,\quad \lambda=-mc^2-k|{\bf j}|,\quad \lambda=-mc^2+k|{\bf j}|.
    \label{rest}
\end{equation}
Equivalently, the dispersion relation (\ref{dispersion}) reduces to $[\lambda^2-(mc^2-k|{\bf j}|)^2][\lambda^2-(mc^2+k|{\bf j}|)^2]=0$, which gives the roots (\ref{rest}).
Consequently, in the rest frame of reference, a free spinor particle has two positive-energy, matter states: $E=mc^2-k|{\bf j}|$ and $E=mc^2+k|{\bf j}|$, and two negative-energy, antimatter states: $E=-mc^2-k|{\bf j}|$ and $E=-mc^2+k|{\bf j}|$.
In a frame of reference, in which a spinor particle has momentum ${\bf p}$, these states have more complicated energies (\ref{energy}), which depend on the helicity ${\bf p}\cdot{\bf j}$.

To reach a lower energy in the rest frame, a particle of matter prefers to be in the state with $E=mc^2-k|{\bf j}|$, corresponding to the effective mass $M=E/c^2=m-k|{\bf j}|/c^2$, and a particle of antimatter prefers to be in the state with $E=-mc^2-k|{\bf j}|$.
Because particles of antimatter can be regarded as positive-energy states under time inversion, the preferred energy state of a time-inversed antimatter particle is $E=mc^2+k|{\bf j}|$, corresponding to the effective mass $M=m+k|{\bf j}|/c^2$.
Consequently, particles of matter and antimatter in the presence of torsion have different effective masses \cite{anti}, which are also different from the spinor mass $m$ that determines the dispersion relation in the absence of torsion.
On the average, particles of antimatter in the presence of torsion have larger masses than the corresponding particles of matter.
Because a spinor has two positive-energy and two negative-energy states, the relations (\ref{rest}) give an inequality $k|{\bf j}|<mc^2$, determining an upper limit on $|{\bf j}|$ and thus on the mass density in the Universe.\\

\noindent
{\bf Four-velocity of a spinor with torsion and wave--particle duality}.\\
In the absence of torsion, the normalized solution of (\ref{plane}) is given by
\begin{equation}
    \psi({\bf r},t)=\left( \begin{array}{cc}
    (E+mc^2)I_2 & {\bm\sigma}\cdot{\bf p}c \\
    {\bm\sigma}\cdot{\bf p}c & (E+mc^2)I_2 \end{array} \right)\left( \begin{array}{c}
    \xi \\
    \eta \end{array} \right)\frac{1}{\sqrt{2mc^2(E+mc^2)}}\exp[i({\bf p}\cdot{\bf r}-Et)/\hbar],
    \label{function}
\end{equation}
where the square matrix is the spinor representation of the boost from rest to the velocity ${\bf v}={\bf p}c^2/E$ \cite{FranciscoNiko,QM} and $\xi$ and $\eta$ are two-dimensional (up and down) spinors describing positive (particle) and negative (antiparticle) energy states.

For a spin-up, positive-energy state, $\xi^\dagger=(1,0)$ and $\eta^\dagger=(0,0)$.
The scalar bilinear composed from the plane wave (\ref{function}) is
\begin{eqnarray}
    & & \bar{\psi}\psi=\psi^\dagger\gamma^0\psi=\frac{1}{2mc^2(E+mc^2)}\Bigl[(E+mc^2)^2-(1,0)({\bm\sigma}\cdot{\bf p})({\bm\sigma}\cdot{\bf p})c^2
    \left( \begin{array}{c}
    1 \\
    0 \\ \end{array} \right)\Bigr] \nonumber \\
    & & =\frac{(E+mc^2)^2-{\bf p}^2 c^2}{2mc^2(E+mc^2)}=1,
\end{eqnarray}
as expected.
The vector bilinear components composed from (\ref{plane}), using the relation $p^\mu=mcu^\mu$ for a free particle and an identity $\sigma^\mu\sigma^\nu+\sigma^\nu\sigma^\mu=-2\eta^{\mu\nu}I_2$, are
\begin{eqnarray}
    & & \bar{\psi}\gamma^0\psi=\psi^\dagger\psi=\frac{1}{2mc^2(E+mc^2)}\Bigl[(E+mc^2)^2+(1,0)({\bm\sigma}\cdot{\bf p})({\bm\sigma}\cdot{\bf p})c^2
    \left( \begin{array}{c}
    1 \\
    0 \\ \end{array} \right)\Bigr] \nonumber \\
    & & =\frac{(E+mc^2)^2+{\bf p}^2 c^2}{2mc^2(E+mc^2)}=\frac{E}{mc^2}=u^0,
\end{eqnarray}
and for the space components:
\begin{eqnarray}
    & & \bar{\psi}\gamma^\mu\psi=\psi^\dagger\alpha^\mu\psi=\frac{E+mc^2}{2mc^2(E+mc^2)}\Bigl[(1,0)\sigma^\mu({\bm\sigma}\cdot{\bf p})c\left( \begin{array}{c}
    1 \\
    0 \\ \end{array} \right)+(1,0)({\bm\sigma}\cdot{\bf p})c\sigma^\mu\left( \begin{array}{c}
    1 \\
    0 \\ \end{array} \right)\Bigr] \nonumber \\
    & & =\frac{p^\mu}{mc}=u^\mu.
\end{eqnarray}
The same result is obtained for a spin-down, positive-energy state, $\xi^\dagger=(0,1)$ and $\eta^\dagger=(0,0)$.
Consequently, the four-velocity of a spinor representing a matter particle is given by \cite{FranciscoNiko}
\begin{equation}
    u^i=\frac{\bar{\psi}\gamma^i\psi}{\bar{\psi}\psi}.
    \label{velocity}
\end{equation}

For a spin-up, negative-energy state, $\xi^\dagger=(0,0)$ and $\eta^\dagger=(1,0)$.
For a spin-down, negative-energy state, $\xi^\dagger=(0,0)$ and $\eta^\dagger=(0,1)$.
Similar calculations for these states give
\begin{equation}
    \bar{\psi}\psi=-1,\quad \bar{\psi}\gamma^0\psi=u^0,\quad \bar{\psi}\gamma^\mu\psi=u^\mu.
\end{equation}
Consequently, the four-velocity of a spinor representing an antimatter particle is given by
\begin{equation}
    u^i=-\frac{\bar{\psi}\gamma^i\psi}{\bar{\psi}\psi}.
    \label{antivelocity}
\end{equation}
The four-velocities (\ref{velocity}) and (\ref{antivelocity}) have a covariant form and satisfy the equations of motion (\ref{motion}) \cite{FranciscoNiko}, so they represent the general-relativistic wave--particle duality.
In the presence of torsion, the mass $m$ in the spinor wave function (\ref{function}) should be replaced by one of the values $\lambda/c^2$ in (\ref{rest}).
Accordingly, the four-velocities (\ref{velocity}) and (\ref{antivelocity}) are also valid in the Einstein--Cartan theory of gravity \cite{FranciscoNiko}.\\

\noindent
{\bf Matter--antimatter asymmetry from torsion}.\\
The energy of a free Dirac particle in the presence of torsion therefore depends on whether it is a matter particle or an antimatter particle.
Accordingly, torsion violates the charge conjugation (C) symmetry.
The energy also depends on helicity, violating the charge conjugation--parity (CP) symmetry.
At densities much smaller than the Cartan density, $k|{\bf j}|\ll mc^2$, so the effects of torsion are negligible and so are these symmetry violations.
At these densities, the masses of corresponding matter and antimatter particles are almost equal, in accordance with the results of experimental measurements of the masses of the proton and antiproton \cite{antiproton}.
The matter--antimatter mass asymmetry, and the corresponding helicity asymmetry, might be responsible for the observed spin asymmetry in the proton \cite{proton}, and could possibly be measured directly in experiments probing smaller distances and higher densities.\\

\noindent
{\bf Torsion-generated asymmetry in gravitational capture by black holes}.\\
The effective cross-section for gravitational capture of a particle with mass $m$, moving with an arbitrary velocity $v_\infty$ at spatial infinity, by a Schwarzschild black hole with mass $\mu$ is determined from the effective potential energy \cite{LL2}
\begin{equation}
    U(r)=mc^2\left[\left(1-\frac{r_g}{r}\right)\left(1+\frac{M^2}{m^2 c^2 r^2}\right)\right]^{1/2},
\end{equation}
where $r_g=2G\mu/c^2$ is the Schwarzschild radius of the black hole and $M$ is the angular momentum of the particle.
In the radial equation of motion, $(1-r_g/r)^{-1}dr/c\,dt=(1/E)(E^2-U^2)^{1/2}$, the condition $E\ge U$ gives the range of admissible motion of the particle.
Its conserved energy is $E=mc^2\gamma$, where $\gamma=(1-v^2_\infty/c^2)^{-1/2}$, and its conserved angular momentum is $M=m\rho v_\infty\gamma$, where $\rho$ is the impact parameter.

The particle falls into the black hole if $E>U_\textrm{max}$, where the maximum value $U_\textrm{max}$ of the effective potential energy is given by $dU/dr=0$ and $d^2U/dr^2<0$.
These conditions give $\rho<\rho_\textrm{max}$, from which the capture cross-section is $\sigma=\pi\rho^2_\textrm{max}$, leading to \cite{cross}
\begin{equation}
    \sigma=\frac{1}{8}\pi r^2_g\left[-(\alpha^2-18\alpha-27)+\sqrt{\alpha^4-28\alpha^3+270\alpha^2+972\alpha+729}\right],
\end{equation}
where $\alpha=1/(\gamma^2-1)$.
For an ultrarelativistic particle, $\alpha\ll 1$ gives
\begin{equation}
    \sigma\approx\frac{27}{4}\pi r^2_g\left(1+\frac{2}{3\gamma^2}\right).
    \label{crosssection}
\end{equation}
For light, this formula reduces to $\sigma=(27/4)\pi r^2_g$.
For a nonrelativistic particle, $\alpha\to\infty$ gives $\sigma\to 4\pi r^2_g c^2/v^2_\infty$ \cite{limit}.

In the very early Universe, strong and time-varying gravitational fields caused intense pair production of fermions \cite{infl,prod}.
In the rest frame of reference of a matter--antimatter fermion pair, the conservation of momentum gives $M_q v_q\gamma_q=M_{\bar{q}} v_{\bar{q}}\gamma_{\bar{q}}$, where the subscript $q$ denotes matter and $\bar{q}$ denotes antimatter.
For ultrarelativistic particles produced in the early Universe, their velocities were $v_q\approx v_{\bar{q}}\approx c$, giving
\begin{equation}
    \gamma_{\bar{q}}/{\gamma_q}\approx M_q/M_{\bar{q}}.
    \label{momentum}
\end{equation}
In the presence of torsion, the effective masses are $M_{-}=m-k|{\bf j}|/c^2$ and $M_{+}=m+k|{\bf j}|/c^2$, following (\ref{rest}).
Particles of matter preferred to be in the state with $M_{-}$ and particles of antimatter preferred to be in the state with $M_{+}$.
The energy difference between the two states is $\Delta E=2k|{\bf j}|$.
In thermodynamical equilibrium, the ratios of the numbers of particles in these states are given by the Boltzmann distribution \cite{LL5}:
\begin{equation}
    \frac{N_{q+}}{N_{q-}}=\frac{N_{\bar{q}-}}{N_{\bar{q}+}}=\exp\left(-\frac{2k|{\bf j}|}{k_\textrm{B}T}\right),
    \label{dist1}
\end{equation}
where $T$ is the temperature of the Universe.
The subscripts $-$ and $+$ denote the states with $M_{-}$ and $M_{+}$, respectively.
These numbers also satisfy
\begin{equation}
    N_{q+}+N_{q-}=N_{\bar{q}-}+N_{\bar{q}+}=N,
    \label{dist2}
\end{equation}
where $N$ is the number of all matter fermions, equal to that of all antimatter fermions because of pair production.

The value of $|{\bf j}|$ can be estimated by normalization of the spinor wave function.
Following the Dirac equation (\ref{plane}), the dimension of the volume integral of $\bar{\psi}\psi$ is the same as that of $c$.
The volume of a closed Universe is $2\pi^2 a^3$, where $a$ is the scale factor.
Therefore, the volume occupied by the wave function of each fermion is approximately $2\pi^2 a^3/2N$, which is equal to the inverse of the number density of all fermions $n=2N/2\pi^2 a^3$.
Consequently, $|{\bf j}|\sim\bar{\psi}\psi$ is given by $\bar{\psi}\psi(2\pi^2 a^3/2N)=c$, which is equivalent to
\begin{equation}
    |{\bf j}|\approx nc.
    \label{density}
\end{equation}
The relations (\ref{dist1}), (\ref{dist2}), and (\ref{density}) determine the numbers of fermion particles:
\begin{equation}
    N_{q+}=N_{\bar{q}-}=\frac{N}{1+e^{2knc/k_\textrm{B}T}},\quad 
    N_{q-}=N_{\bar{q}+}=\frac{Ne^{2knc/k_\textrm{B}T}}{1+e^{2knc/k_\textrm{B}T}}.
    \label{dist3}
\end{equation}

The mean effective cross-sections for gravitational capture of fermions are given by (\ref{crosssection}) and (\ref{dist3}):
\begin{equation}
    \sigma_q=\frac{27}{4N}\pi r^2_g\left[\left(1+\frac{2}{3\gamma^2_{-}}\right)N_{q-}+\left(1+\frac{2}{3\gamma^2_{+}}\right)N_{q+}\right],\quad \sigma_{\bar{q}}=\frac{27}{4N}\pi r^2_g\left[\left(1+\frac{2}{3\gamma^2_{-}}\right)N_{\bar{q}-}+\left(1+\frac{2}{3\gamma^2_{+}}\right)N_{\bar{q}+}\right].
\end{equation}
The relative difference of the cross-sections for matter and antimatter particles is therefore equal to
\begin{equation}
    \frac{\Delta\sigma}{\sigma_q}=\frac{\sigma_{\bar{q}}-\sigma_q}{\sigma_q}=\frac{2}{3}\frac{(e^{2knc/k_\textrm{B}T}-1)(1/\gamma^2_{+}-1/\gamma^2_{-})}{(1+2/3\gamma^2_{-})e^{2knc/k_\textrm{B}T}+(1+2/3\gamma^2_{+})},
\end{equation}
which for ultrarelativistic particles with $\gamma\gg1$ reduces (in the denominator) to
\begin{equation}
    \frac{\Delta\sigma}{\sigma_q}\approx\frac{2}{3}\frac{e^{2knc/k_\textrm{B}T}-1}{e^{2knc/k_\textrm{B}T}+1}\left(\frac{1}{\gamma^2_{+}}-\frac{1}{\gamma^2_{-}}\right)=\frac{2}{3\gamma^2_{-}}\tanh(knc/k_\textrm{B}T)\left(\frac{\gamma^2_{-}}{\gamma^2_{+}}-1\right).
    \label{difference}
\end{equation}
This relation can be simplified, using $\gamma_{-}/\gamma_{+}\approx M_{+}/M_{-}$ in the last term, which follows from (\ref{momentum}) and the conservation of helicity in pair production, and $M_{+}\approx M_{-}$:
\begin{equation}
    \frac{\gamma^2_{-}}{\gamma^2_{+}}-1\approx \frac{M^2_{+}}{M^2_{-}}-1\approx\frac{2\Delta M}{M_{-}},
\end{equation}
where $\Delta M=M_{+}-M_{-}=2k|{\bf j}|/c^2\approx 2kn/c$ is the effective mass difference and $M_{-}\approx m$.
The mean value of $\gamma^2_{-}$ in (\ref{difference}) can be determined from the relativistic equipartition theorem, which gives $mc^2\gamma=3k_\textrm{B}T$, so $\gamma^2_{-}\approx 9(k_\textrm{B}T/mc^2)^2$.
Therefore, the relative cross-section difference is
\begin{equation}
    \frac{\Delta\sigma}{\sigma_q}\approx\frac{8kn}{3m\gamma^2_{-}c}\tanh(knc/k_\textrm{B}T)\approx \frac{8kmnc^3}{27(k_\textrm{B}T)^2}\tanh(knc/k_\textrm{B}T).
    \label{semifinal}
\end{equation}

In thermodynamical equilibrium, the fermion number density is given by \cite{LL5}
\begin{equation}
    n=\frac{\zeta(3)}{\pi^2}\frac{3g}{4}\frac{(k_\textrm{B}T)^3}{(\hbar c)^3},
    \label{numberdensity}
\end{equation}
where $g=90$ is the number of thermal degrees of freedom for all known elementary fermions.
Using this formula and also the definitions of the Planck mass $m_\textrm{P}=\sqrt{\hbar c/G}$ and the Planck temperature $T_\textrm{P}=\sqrt{\hbar c^5/Gk_\textrm{B}^2}$, gives the formula for the relative cross-section difference between matter and antimatter in terms of the original mass $m$ of a particle (in the absence of torsion) and the temperature $T$:
\begin{equation}
    \frac{\Delta\sigma}{\sigma_q}\approx\frac{2\zeta(3)g}{3\pi}\frac{m}{m_\textrm{P}}\frac{T}{T_\textrm{P}}\tanh\left(\frac{9\zeta(3)g}{4\pi}\frac{T^2}{T^2_\textrm{P}}\right)\approx\frac{3\zeta^2(3)g^2}{2\pi}\frac{m}{m_\textrm{P}}\frac{T^3}{T^3_\textrm{P}},
    \label{final}
\end{equation}
using $\tanh(x)\approx x$ for $T\ll T_\textrm{P}$.
Even for $T\sim T_\textrm{P}$, this quantity is very low because $m\ll m_\textrm{P}$.\\

\noindent
{\bf Baryogenesis and leptogenesis from primordial black holes and torsion}.\\
The mass difference between a fermion and the corresponding antifermion increases with density and becomes significant near the Cartan density, which existed in the early Universe.
Because antimatter particles were more massive than matter particles, during pair production in the early Universe the antimatter particles were slower than the corresponding matter particles.
Because the cross-section for gravitational capture of a massive particle by a black hole increases as its speed decreases, the antimatter particles have a larger capture cross section than the corresponding matter particles.

If the early Universe produced many primordial black holes \cite{primordial}, more antimatter particles than matter particles fell into them.
We propose that the torsion-generated mass asymmetry and the resulting black-hole capture asymmetry produced the matter--antimatter imbalance in the observable Universe: the missing antimatter fell into black holes and the remaining antimatter annihilated with matter, producing photons and a surplus of matter.
This mechanism does not violate the conservation of baryon number and thus does not need new physics beyond the Standard Model of particle physics \cite{capture}.

The torsion-generated baryogenesis satisfies the three Sakharov conditions to produce matter and antimatter at different rates \cite{Sak}.
Baryon number violation in the Universe effectively follows from antimatter falling to black holes at higher rates than matter.
C-symmetry violation and CP-symmetry violation follow from the mass depending on whether the particle composes matter or antimatter and on helicity.
Interactions out of thermal equilibrium follow from the unidirectional motion of particles through the event horizon of a black hole.

Because almost all antimatter in the Universe, which was not captured by primordial black holes, annihilated with matter and produced photons, the quantity (\ref{final}) describes the baryon-to-photon ratio $\eta$ in the Universe.
This quantity is very low, which explains the enormous abundance of photons over matter. 
However, it is much lower than the value of the baryon-to-photon ratio $\eta$, which is measured at about $6.1\times 10^{-10}$ \cite{photon}.
However, the coupling between spin and torsion generates a negative correction to the energy density, which acts like gravitational repulsion.
Pair production in the early Universe, caused by strong and time-varying gravitational fields \cite{prod}, increased this repulsion, giving a finite period of exponential expansion of the Universe and providing the simplest explanation of cosmic inflation \cite{infl}.
Pair production violated thermal equilibrium, so the relation $n\sim T^3$ (\ref{numberdensity}) was not satisfied.
Consequently, the value of $n$ in (\ref{semifinal}) is increased by many orders of magnitude.
Also, the fraction of particles that fell into primordial black holes is proportional to the fraction of space occupied by them at the time of pair production.
This fraction decreases the value of $n$ in (\ref{semifinal}).
These two factors must be taken into account before comparing the value (\ref{final}) to $\eta$.

The proposed explanation of the observed matter--antimatter asymmetry in the Universe is based on spacetime torsion.
In addition to this asymmetry and cosmic inflation, torsion may solve several other problems in general relativity and quantum field theory.
Torsion imposes a spatial extension on fermions \cite{non} and removes the ultraviolet divergence of radiative corrections represented by loop Feynman diagrams \cite{torsional}.
The gravitational repulsion produced by the coupling between spin and torsion in the Einstein--Cartan theory of gravity might have prevented the formation of gravitational singularities at the beginning of the Universe \cite{avert,universe}, preserving the inequality $k|{\bf j}|<mc^2$ and thus determining an upper limit on the mass density in the Universe.
This result is also valid for spacetime with shear \cite{shear}.
This repulsion may also prevent the formation of singularities in black holes.
Consequently, the collapsing matter in a black hole would avoid a singularity and instead reach a nonsingular bounce, after which it would expand as a new, closed universe on the other side of its event horizon, with a finite period of inflation \cite{infl}.
Accordingly, our Universe might have originated as a baby universe in a black hole existing in another, parent universe \cite{Pat}, with the Big Bang being actually a Big Bounce.
Torsion provides the simplest and most natural explanation of this origin \cite{infl,universe}.

I am grateful to Francisco Guedes and my Parents, Bo\.{z}enna Pop{\l}awska and Janusz Pop{\l}awski, for supporting this work.

\end{document}